\documentclass[aps,prl,comment,twocolumn,superscriptaddress,groupedaddress]{revtex4}  
\usepackage{graphicx}  
\usepackage{dcolumn}   
\usepackage{bm}        
\usepackage{amssymb}   

\begin{document}

\title{Comment on ``Axion Dark Matter Coupling to Resonant Photons via Magnetic Field"}
\date{June 29, 2016}
\author{Sangjun Lee} \affiliation{Dept. of physics, KAIST, Daejeon 34141 Republic of Korea} \affiliation{Center for Axion and Precision Physics Research, IBS, Daejeon 34141 Republic of Korea}
\author{Sung Woo Youn} \affiliation{Center for Axion and Precision Physics Research, IBS, Daejeon 34141 Republic of Korea}
\author{Y. K. Semertzidis} \affiliation{Dept. of physics, KAIST, Daejeon 34141 Republic of Korea} \affiliation{Center for Axion and Precision Physics Research, IBS, Daejeon 34141 Republic of Korea}

\maketitle

A recent Letter~\cite{bib:report} claims that for typical dark matter axion search experiments using cylindrical haloscopes, the power gain depends on the relative position of a cavity with respect to the center of a solenoidal magnetic field due to different electric and magnetic couplings. We review this Letter and find a misinterpretation of the coordinate system. We correct this and see no dependence of the coupling strength on the cavity location and the electric and magnetic energies stored in a cavity mode are equal.

From the modified Maxwell's equations in the presence of axion-photon coupling with the background axion field $a=a_0 e^{-i\omega_a t}$, the electromagnetic field components of photons produced via the axion-photon conversion inside the solenoid producing homogeneous magnetic field $\vec{B_0}=B_0\hat{z}$ are obtained as
\begin{equation}
\vec{E}_a=-g_{a\gamma\gamma} acB_0 \hat{z}, \qquad
 \vec{B}_a= -\frac{g_{a\gamma\gamma}}{2c}rB_0\frac{\partial{a}}{\partial{t}}\hat{\phi},
\end{equation}
where $g_{a\gamma \gamma}$ is the axion-photon coupling constant and $(r, \phi)$ is the polar coordinate system of the solenoid. Since $\vec{E}_a$ and $\vec{B}_a$ are determined by the boundary condition of the solenoid, they do not depend on the location of the cavity.

Denoting by $\vec{E}_c$ and $\vec{B}_c$ the electric and magnetic fields of a resonant cavity mode, the electric and magnetic energies stored in the cavity mode from the axion-to-photon conversion are given by
\begin{eqnarray}
U_{a,e}&=&\frac{1}{4}\epsilon_0 g_{a\gamma\gamma}^2 a^2c^2 B_0^2 V C_E,\\
U_{a,m}&=&\frac{1}{4}\epsilon_0 g_{a\gamma\gamma}^2 a^2c^2 B_0^2 V C_B,
\end{eqnarray}
with the electric and magnetic form factors being defined, respectively, as 
\begin{equation}
C_E = \frac{\left| \int{dV_c}~\vec{E}_c \cdot \hat{z} \right|^2}{V\int{dV_c}\left| E_c \right|^2},\ \ \
C_B = \frac{\frac{\omega_a^2}{c^2}\left| \int{dV_c}~\frac{r}{2} \vec{B}_c \cdot \hat{\phi} \right|^2}{V\int{dV_c}\left| B_c \right|^2},
\label{form_factor}
\end{equation}
where the integration is over the cavity volume.

At a first glance, $C_B$ in Eq.~\ref{form_factor} would produce different results when the cavity is no longer coaxial with the solenoid and has some offset $e$ due to the apparent radial dependence ($C_E$ remains unchanged since $\vec{E}_a$ is homogeneous inside the solenoid). However, in the presence of the offset, $\vec{B}_c$ lies in the azimuthal direction in the cavity coordinates, $\hat{\phi}_c$, while $r$ and $\hat{\phi}$ in the integrand of $C_B$ remain the same in the solenoid coordinates since these are due to the $\vec{B}_a$ field. Thus we evaluate the dot product between $\hat{\phi}_c$ and $\hat{\phi}$, i.e., $\hat{\phi}_c \cdot \hat{\phi}=\cos(\phi+\phi_c) = -[r_c -e\cos\phi_c]/r$. Then $C_B$ in Eq.~\ref{form_factor} becomes
\begin{equation}
C_B = \frac{\frac{\omega_a ^2}{c^2} \left| \int{dV_c}~B_{c_{\phi_c}}\frac{r_c-e\cos\phi_c}{2} \right|^2 } {V\int{dV_c}\left| B_c \right|^2 }.
\label{C_B}
\end{equation}
Recalling the corresponding expression in the Letter,
\begin{equation}
C_B = \frac{\frac{\omega_a ^2}{c^2} \left| \int{dV_c}~B_{c_{\phi}}\frac{r-e\cos\phi}{2} \right|^2 } {V\int{dV_c}\left| B_c \right|^2 },
\end{equation}
we find the radial and azimuthal coordinates are interpreted in the solenoid system rather than the cavity system, which in turn results in $C_B$ varying with the offset.
It must be emphasized that $r_c$ and $\phi_c$ are the polar coordinates of the cavity.
Now the $\cos\phi_c$ term in the integrand of Eq.~\ref{C_B} vanishes after integration over the cavity volume. Therefore, it is found that $C_B$ has no position (offset $e$) dependence in the solenoid and has the same value as $C_E$, e.g., $C_B = C_E = 0.69$ for the $\mathrm{TM}_{010}$ mode.

This work was supported by IBS-R017-D1-2016-a00/IBS-R017-Y1-2016-a00. We also thank S. H.~Chang at the Center for Theoretical Physics of the Universe (CTPU) of Institute for Basic Science (IBS) for his confirmation of our finding.


\begin{thebibliography}{99}

\bibitem{bib:report} B. T. McAllister, S. R. Parker, and M. E. Tobar, ``Axion Dark Matter Coupling to Resonant Photons via Magnetic Field", Phys.\ Rev.\ Lett.\ {\bf 116}, 161804 (2016). 
\end{thebibliography}
\end{document}